\documentclass[11pt]{article}

\usepackage{PhysRep}
\usepackage{Lep2Rep}
\usepackage{gg}
\usepackage{ff}
\usepackage{4f_s06}
\usepackage{smat}
\usepackage{gc}
\usepackage{mw}

\usepackage[english]{babel}
\usepackage{graphicx,rotating}
\usepackage{a4p,here}
\usepackage{cite,mcite,epsfig}

\renewcommand{\Huge}{\huge}
\newcommand{\updates}[1]%
{\fbox{\parbox{\linewidth}{\textbf{Updates with respect to summer 2007:}\\#1}}}

\input{rotate}

\begin{document}
\begin{titlepage}
\begin{center}
\Large {EUROPEAN ORGANIZATION FOR NUCLEAR RESEARCH\\
FERMI NATIONAL ACCELERATOR LABORATORY\\
STANFORD LINEAR ACCELERATOR CENTER}
\end{center}

\begin{flushright}
       CERN-PH-EP-2010-095 \\
       FERMILAB-TM-2480-PPD \\ %
       LEPEWWG/2010-01 \\
       TEVEWWG/2010-01 \\
       ALEPH 2010-001 PHYSICS 2010-001 \\
       CDF Note 10338 \\
       D0 Note 6130 \\
       DELPHI DELPHI-2010-001 PHYS 952 \\
       L3 Note 2837 \\
       OPAL PR432 \\
       SLAC-PUB-14301\\
       {\bf 10-DEC-2010 }
\end{flushright}

\begin{center}
\boldmath
\Huge {\bf %
           Precision 
           Electroweak Measurements and \\
           Constraints on the Standard Model\\[.5cm]

}
\unboldmath

\vspace*{1.0cm}
\Large {\bf
The 
ALEPH, CDF, D0, DELPHI, L3, OPAL, SLD Collaborations,\\
the LEP Electroweak Working Group,\footnote{ WWW access at {\tt http://www.cern.ch/LEPEWWG}}\\
the Tevatron Electroweak Working Group,\footnote{ WWW access at {\tt http://tevewwg.fnal.gov}}\\
and the SLD electroweak and heavy flavour groups}

\vskip 0.5cm
\large\textbf{Prepared from Contributions %
to the 2010 Summer Conferences.}\\
\end{center}
\vfill
\begin{abstract}

This note presents constraints on Standard Model parameters using
published and preliminary precision electroweak results obtained at
the electron-positron colliders LEP and SLC.
  The results are compared with precise electroweak measurements from
  other experiments, notably CDF and D\O\ at the Tevatron.
  Constraints on the input parameters of the Standard Model are
  derived from the combined set of results obtained in high-$Q^2$
  interactions, and used to predict results in low-$Q^2$ experiments,
  such as atomic parity violation, M{\o}ller scattering, and
  neutrino-nucleon scattering. The main changes with respect to the
  experimental results presented in 2009 are new combinations of
  results on the width of the W boson and the mass of the top quark.

\end{abstract}
\vfill
\end{titlepage}
\setcounter{page}{2}
\renewcommand{\thefootnote}{\arabic{footnote}}
\setcounter{footnote}{0}

\boldmath
\unboldmath

\section{Introduction}

The experimental results used here consist of the final and published
Z-pole results~\cite{bib-Z-pole} measured by the ALEPH, DELPHI, L3,
OPAL and SLC experiments, taking data at the electron-positron
colliders LEP and SLC. In addition, published and preliminary results
on the mass and width of the W boson, measured at {\LEPII} and the
Tevatron, and the mass of the top quark, measured at the Tevatron
only, are included. This report updates our previous analysis from
2009~\cite{bib-EWEP-09}; the main changes with respect to the
experimental results presented there are new combinations of results
on the width of the W boson and the mass of the top quark.

The measurements %
allow to check the validity of the Standard Model (SM) and, within its
framework, to infer valuable information about its fundamental
parameters. The accuracy of the W- and Z-boson measurements makes them
sensitive to the mass of the top quark, $\Mt$, and to the mass of the
Higgs boson, $\MH$, through loop corrections. While the leading $\Mt$
dependence is quadratic, the leading $\MH$ dependence is logarithmic.
Therefore, the inferred constraints on $\Mt$ are much stronger than
those on $\MH$.  Independent analyses of this type are presented in
References~\citen{Gfitter} and~\citen{common_bib:pdg2010}, obtaining
equivalent results when accounting for different measurements used as
input.

\section{Measurements}

The measurements considered here are reported in Table~\ref{tab-SMIN}.
Also shown are the predictions based on the results of the SM fit to
these high-$Q^2$ measurements, as reported in the last column of
Table~\ref{tab-BIGFIT}.

The measurements obtained at the Z pole by the LEP and SLC experiments
ALEPH, DELPHI, L3, OPAL and SLD, and their combinations, reported in
parts a), b) and c) of Table~\ref{tab-SMIN}, are final and
published~\cite{bib-Z-pole}.

The results on the W-boson mass by CDF~\cite{CDF-MW-PRL90,
  *CDF-MW-PRD90, *CDF-MW-PRL95, *CDF-MW-PRD95, *CDF-MW-2000} and
D\O~\cite{D0-MW:central, *D0-MW:endcap, *D0-MW:edge, *D0-MW:large} in
Run~I, and on the W-boson width by CDF\cite{CDF-GW} and
D\O\cite{D0-GW} in Run~I, are combined by the Tevatron Electroweak
Working Group based on a detailed treatment of common systematic
uncertainties~\cite{PP-MW-GW:combination}.  Including also results
based on Run~II data on $\MW$ by CDF~\cite{CDF2MWPRL,CDF2MWPRD} and
D\O~\cite{D02MW}, and on $\GW$ by CDF~\cite{CDF2GW} and
D\O~\cite{D02GW}, the combined Tevatron results
are~\cite{PP-MW:2009,PP-GW:2010}: $\MW = 80.420\pm0.031~\GeV$, $\GW =
2.046\pm0.049~\GeV$.  Including the preliminary {\LEPII}
combination~\cite{bib-EWEP-06}, $\MW=80.376\pm0.033~\GeV$ and
$\GW=2.196\pm0.083~\GeV$,
the resulting averages are:
\begin{eqnarray}
\MW & = & 80.399 \pm 0.023~\GeV\\
\GW & = &  2.085 \pm 0.042~\GeV\,.
\end{eqnarray}

For the mass of the top quark, $\Mt$, the published Run~I results from
CDF~\cite{Mtop1-CDF-di-l-PRLa, *Mtop1-CDF-di-l-PRLb,
*Mtop1-CDF-di-l-PRLb-E, *Mtop1-CDF-l+j-PRL, *Mtop1-CDF-l+j-PRD,
*Mtop1-CDF-all-j-PRL} and D\O~\cite{D0-top:prl-ll, *D0-top:prd-ll,
*D0-top:prl-lj, *D0-top:prd-lj, *Mtop1-D0-l+j-new1}, and preliminary
and published results based on Run~II data from
CDF~\cite{Mtop2-CDF-all-j-new, *Mtop2-CDF-trk-new,
*Mtop2-CDF-di-l-new, *Mtop2-CDF-l+j-new} and
D\O~\cite{Mtop2-D0-l+j-new, *Mtop2-D0-di-l-mar09-1,
*Mtop2-D0-di-l-jul08-2}, are combined by the Tevatron Electroweak
Working Group with the result:
$\Mt=173.3\pm1.1~\GeV$~\cite{TeVEWWGtop-1007} where the uncertainty
includes an estimate for colour reconnection effects. The exact
definition of the mass of the top quark and the related theoretical
uncertainty in the interpretation of the measured ``pole'' mass
require further study~\cite{LMUtop:2009}.

In addition, the following results obtained in low-$Q^2$ interactions
and reported in Table~\ref{tab-SMpred} are considered: (i) the
measurements of atomic parity violation in caesium\cite{QWCs:exp:1,
QWCs:exp:2}, with the numerical result\cite{QWCs:theo:2003:new} based
on a revised analysis of QED radiative corrections applied to the raw
measurement; (ii) the result of the E-158 collaboration on the
electroweak mixing angle\footnote{ E-158 quotes the experimental
result in the MSbar scheme, evolved to $Q^2=\MZ^2$.  We add 0.00029 to
the quoted value in order to obtain the effective electroweak mixing
angle~\cite{common_bib:pdg2010}.}  measured in M{\o}ller
scattering~\cite{E158RunI, *E158RunI+II+III}; and (iii) the final
result of the NuTeV collaboration on neutrino-nucleon
neutral-to-charged current cross section
ratios~\cite{bib-NuTeV-final}.

Using neutrino-nucleon data with an average $Q^2\simeq20~\GeV^2$, the
NuTeV collaboration has extracted the left- and right-handed couplings
combinations 
$\gnlq^2=4\gln^2(\glu^2+\gld^2) =
[1/2-\swsqeff+(5/9)\swsqsqeff]\rhon\rho_{\mathrm{ud}}$ and
$\gnrq^2=4\gln^2(\gru^2+\grd^2) =
(5/9)\swsqsqeff\rhon\rho_{\mathrm{ud}}$, with the $\rho$ parameters
defined in \cite{bib-PCLI}.  The NuTeV results for the effective
couplings are: $\gnlq^2=0.30005\pm0.00137$ and
$\gnrq^2=0.03076\pm0.00110$, with a correlation of $-0.017$.  While
the result on $\gnrq$ agrees with the $\SM$ expectation, the result on
$\gnlq$, measured nearly eight times more precisely than $\gnrq$,
shows a deficit with respect to the expectation at the level of 2.9
standard deviations~\cite{bib-NuTeV-final}.\footnote{A new study finds
that EMC-like isovector effects are able to explain this
difference~\cite{CBT:2009, *BCLT:2009}.}

An additional input parameter, not listed in the table, is the Fermi
constant, $G_F$, determined from the $\mu$ lifetime, $G_F = 1.16637(1)
\cdot 10^{-5}~\GeV^{-2}$\cite{bib-Gmu-1, *bib-Gmu-2, *bib-Gmu-3}.  New
measurements of $G_F$ yield values which are in good
agreement~\cite{Chitwood:2007pa,Barczyk:2007hp}.  The relative error
on $G_F$ is comparable to that of $\MZ$; both errors have negligible
effects on the fit results.

\section{Theoretical and Parametric Uncertainties}

Detailed studies of the theoretical uncertainties in the SM
predictions due to missing higher-order electroweak corrections and
their interplay with QCD corrections were carried out by the working
group on `Precision calculations for the $\Zzero$
resonance'\cite{bib-PCLI}, and later in References~\citen{BP:98}
and~\citen{PCP99}.  Theoretical uncertainties are evaluated by
comparing different but equivalent treatments of aspects such as
resummation techniques, momentum transfer scales for vertex
corrections and factorisation schemes.  The effects of these
theoretical uncertainties are reduced by the inclusion of higher-order
corrections\cite{bib-twoloop,bib-QCDEW} in the electroweak libraries
TOPAZ0~\cite{Montagna:1993py, *Montagna:1993ai, *Montagna:1996ja,
*Montagna:1998kp} and ZFITTER~\cite{Bardin:1989di, *Bardin:1990tq,
*Bardin:1991fu, *Bardin:1991de, *Bardin:1992jc, *Bardin:1999yd,
*Kobel:2000aw, *Arbuzov:2005ma, *ZFITTER643}.

The use of the higher-order QCD corrections increases the value of
$\alfmz$ by 0.001, as expected~\cite{bib-QCDEW}.  The effect of
missing higher-order QCD corrections on $\alfmz$ dominates missing
higher-order electroweak corrections and uncertainties in the
interplay of electroweak and QCD corrections. A discussion of
theoretical uncertainties in the determination of $\alfas$ can be
found in References~\citen{bib-PCLI} and~\citen{bib-SMALFAS}, with a
more recent analysis in Reference~\citen{Stenzel:2005sg} where the
theoretical uncertainty is estimated to be about 0.001 for the
analyses presented here.

The complete (fermionic and bosonic) two-loop corrections for the
calculation of $\MW$~\cite{Twoloop-MW}, and the complete fermionic
two-loop corrections for the calculation of
$\swsqeffl$~\cite{Twoloop-sin2teff} have been calculated.  Including
three-loop top-quark contributions to the $\rho$ parameter in the
limit of large $\Mt$~\cite{Threeloop-rho}, efficient routines for
evaluating these corrections have been implemented since version 6.40
in the semi-analytical program ZFITTER.  The remaining theoretical
uncertainties are estimated to be $4~\MeV$ on $\MW$ and 0.000049 on
$\swsqeffl$.  The latter uncertainty dominates the theoretical
uncertainty in the SM fits and the extraction of constraints on the
mass of the Higgs boson presented below. For a consistent treatment,
the complete two-loop corrections for the partial Z decay widths
should be calculated.

The theoretical uncertainties discussed above are not included in the
results presented in Tables~\ref{tab-BIGFIT} and~\ref{tab-SMpred}.  At
present the impact of theoretical uncertainties on the determination
of $\SM$ parameters from the precise electroweak measurements is small
compared to the impact of the uncertainty on the value of
$\alpha(\MZ^2)$, which is included in the results.

The uncertainty in $\alpha(\MZ^2)$ arises from the contribution of
light quarks to the photon vacuum polarisation, $\dalhad$:
\begin{equation}
\alpha(\MZ^2) = \frac{\alpha(0)}%
   {1 - \Delta\alpha_\ell(\MZ^2) -
   \dalhad - 
   \Delta\alpha_{\mathrm{top}}(\MZ^2)} \,,
\end{equation}
where $\alpha(0)=1/137.036$.  The top contribution, $-0.00007(1)$,
depends on the mass of the top quark, and is therefore determined
inside the electroweak libraries TOPAZ0 and ZFITTER.  The leptonic
contribution is calculated to third order\cite{bib-alphalept} to be
$0.03150$, with negligible uncertainty.

For the hadronic contribution $\dalhad$, we
use the result $0.02758\pm0.00035$~\cite{bib-BP05} which takes into
account published results on electron-positron annihilations into
hadrons at low centre-of-mass energies by the BES
collaboration~\cite{BES_01}, as well as the revised published results
from CMD-2~\cite{CMD_03} and results from KLOE~\cite{KLOE_04}.  The
reduced uncertainty still causes an error of 0.00013 on the $\SM$
prediction of $\swsqeffl$, and errors of 0.2~\GeV{} and 0.1 on the
fitted values of $\Mt$ and $\log(\MH)$, included in the results
presented below.  The effect on the $\SM$ prediction for $\Gll$ is
negligible.  The $\alfmz$ values from the $\SM$ fits presented here
are stable against a variation of $\alpha(\MZ^2)$ in the interval
quoted.

There are also several evaluations of $\dalhad$~\cite{bib-Swartz,
bib-Zeppe, bib-Alemany, bib-Davier, bib-alphaKuhn, bib-jeger99,
bib-Erler, bib-ADMartin, bib-Troconiz-Yndurain, bib-Hagiwara:2003,
bib-Troconiz-Yndurain-2004} which are more theory-driven.  One of the
more recent results~\cite{bib-Troconiz-Yndurain-2004} also includes
the new results from BES, yielding $0.02749\pm0.00012$.  To show the
effects of the uncertainty of $\alpha(\MZ^2)$, we also use this
evaluation of the hadronic vacuum polarisation.
\begin{table}[p]
\begin{center}
\renewcommand{\arraystretch}{1.10}
\begin{tabular}{|ll||r|r|r|r|}
\hline
 && \mcc{Measurement with}  &\mcc{Systematic} & \mcc{Standard} & \mcc{Pull} \\
 && \mcc{Total Error}       &\mcc{Error}      & \mcc{Model fit}&            \\
\hline
\hline
&&&&& \\[-3mm]
& $\dalhad$\cite{bib-BP05}
                & $0.02758 \pm 0.00035$ & 0.00034 &0.02768& $-0.3$ \\
&&&&& \\[-3mm]
\hline
a) & \underline{\LEPI}   &&&& \\
   & line-shape and      &&&& \\
   & lepton asymmetries: &&&& \\
&$\MZ$ [\GeV{}] & $91.1875\pm0.0021\pz$
                & ${}^{(a)}$0.0017$\pz$ &91.1874$\pz$ & $ 0.0$ \\
&$\GZ$ [\GeV{}] & $2.4952 \pm0.0023\pz$
                & ${}^{(a)}$0.0012$\pz$ & 2.4959$\pz$ & $-0.3$ \\
&$\shad$ [nb]   & $41.540 \pm0.037\pzz$ 
                & ${}^{(b)}$0.028$\pzz$ &41.478$\pzz$ & $ 1.7$ \\
&$\Rl$          & $20.767 \pm0.025\pzz$ 
                & ${}^{(b)}$0.007$\pzz$ &20.742$\pzz$ & $ 1.0$ \\
&$\Afbzl$       & $0.0171 \pm0.0010\pz$ 
                & ${}^{(b)}$0.0003\pz & 0.0164\pz     & $ 0.7$ \\
&+ correlation matrix~\cite{bib-Z-pole} &&&& \\
&                                             &&&& \\[-3mm]
&$\tau$ polarisation:                         &&&& \\
&$\cAl~(\ptau)$ & $0.1465\pm 0.0033\pz$ 
                & 0.0016$\pz$ & 0.1481$\pz$ & $-0.5$ \\
                      &                       &&&& \\[-3mm]
&$\qq$ charge asymmetry:                      &&&& \\
&$\swsqeffl(\Qfbhad)$
                & $0.2324\pm0.0012\pz$ 
                & 0.0010$\pz$ & 0.23139     & $ 0.8$ \\
&                                             &&&& \\[-3mm]
\hline
b) & \underline{SLD} &&&& \\
&$\cAl$ (SLD)   & $0.1513\pm 0.0021\pz$ 
                & 0.0010$\pz$ & 0.1481$\pz$ & $ 1.6$ \\
&&&&& \\[-3mm]
\hline
c) & \underline{{\LEPI}/SLD Heavy Flavour} &&&& \\
&$\Rbz{}$        & $0.21629\pm0.00066$  
                 & 0.00050     & 0.21579     & $ 0.8$ \\
&$\Rcz{}$        & $0.1721\pm0.0030\pz$
                 & 0.0019$\pz$ & 0.1723$\pz$ & $-0.1$ \\
&$\Afbzb{}$      & $0.0992\pm0.0016\pz$
                 & 0.0007$\pz$ & 0.1038$\pz$ & $-2.9$ \\
&$\Afbzc{}$      & $0.0707\pm0.0035\pz$
                 & 0.0017$\pz$ & 0.0742$\pz$ & $-1.0$ \\
&$\cAb$          & $0.923\pm 0.020\pzz$
                 & 0.013$\pzz$ & 0.935$\pzz$ & $-0.6$ \\
&$\cAc$          & $0.670\pm 0.027\pzz$
                 & 0.015$\pzz$ & 0.668$\pzz$ & $ 0.1$ \\
&+ correlation matrix~\cite{bib-Z-pole} &&&& \\
&                                              &&&& \\[-3mm]
\hline
d) & \underline{{\LEPII} and Tevatron} &&&& \\
&$\MW$ [\GeV{}] ({\LEPII}, Tevatron)
& $80.399 \pm 0.023\pzz$ &      $\pzz$   & 80.379$\pzz$ & $ 0.9$ \\
&$\GW$ [\GeV{}] ({\LEPII}, Tevatron)
& $ 2.085 \pm 0.042\pzz$ &      $\pzz$   &  2.092$\pzz$ & $ 0.2$ \\
&$\Mt$ [\GeV{}] (Tevatron~\cite{TeVEWWGtop-1007})
& $173.3\pm 1.1\pzz\pzz$ & 0.9$\pzz\pzz$ & 173.4$\pzz\pzz$ & $-0.1$ \\
\hline
\end{tabular}\end{center}
\caption[]{ Summary of high-$Q^2$ measurements included in the
  combined analysis of SM parameters. Section~a) summarises {\LEPI}
  averages, Section~b) SLD results ($\cAl$ includes $\ALR$ and
  the polarised lepton asymmetries), Section~c) the {\LEPI} and SLD
  heavy flavour results, and Section~d) electroweak measurements from
  {\LEPII} and the Tevatron.  The total errors in column 2 include the
  systematic errors listed in column 3.  Although the systematic
  errors include both correlated and uncorrelated sources, the
  determination of the systematic part of each error is approximate.
  The $\SM$ results in column~4 and the pulls (difference between
  measurement and fit in units of the total measurement error) in
  column~5 are derived from the SM fit including all high-$Q^2$ data
  (Table~\ref{tab-BIGFIT}, column~4).\\ $^{(a)}$\small{The systematic
  errors on $\MZ$ and $\GZ$ contain the errors arising from the
  uncertainties in the $\LEPI$ beam energy only.}\\
  $^{(b)}$\small{Only common systematic errors are indicated.}\\ }
\label{tab-SMIN}
\end{table}

\section{Selected Results}

\begin{figure}[htbp]
\begin{center}
$ $ \vskip -1cm
  \mbox{\includegraphics[width=0.9\linewidth]{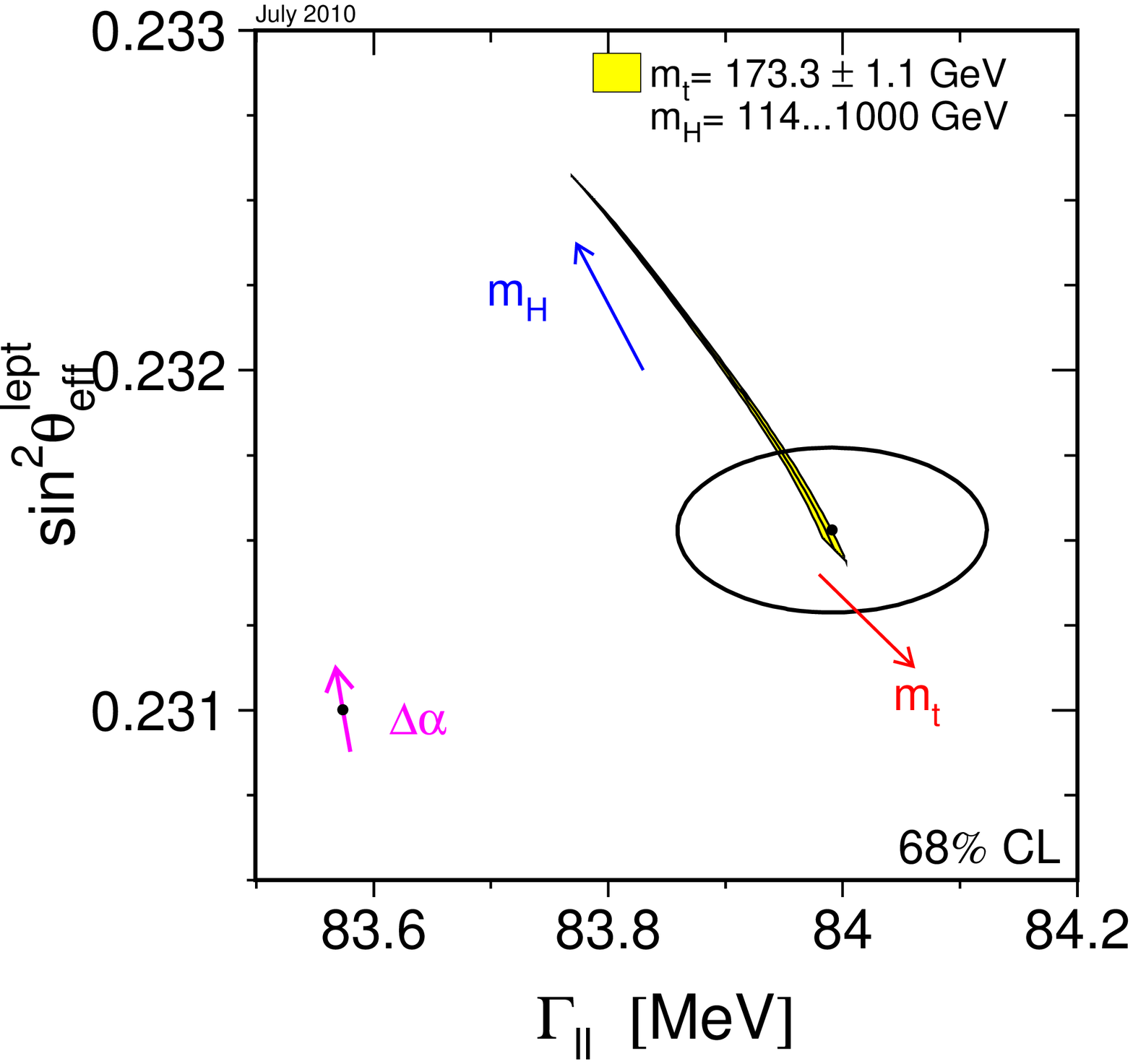}}
\end{center}
\vskip -1cm
\caption[]{ $\LEPI$+SLD measurements~\cite{bib-Z-pole} of $\swsqeffl$
  and $\Gll$ and the SM prediction.  The point with the arrow labelled
  $\Delta\alpha$ shows the prediction when only the photon vacuum
  polarisation is included in the electroweak radiative corrections.
  The associated arrow shows the variation in this prediction if
  $\alpha(\MZ^2)$ is changed by one standard deviation. This variation
  gives an additional uncertainty to the SM prediction shown in the
  figure.  }
\label{fig-gllsef}
\end{figure}

Figure~\ref{fig-gllsef} shows a comparison of the leptonic partial
width from {\LEPI}, $\Gll=83.985\pm0.086~\MeV$~\cite{bib-Z-pole}, and
the effective electroweak mixing angle from asymmetries measured at
{\LEPI} and SLD, $\swsqeffl=0.23153\pm0.00016$~\cite{bib-Z-pole}, with
the SM shown as a function of $\Mt$ and $\MH$.  Good agreement with
the $\SM$ prediction using the most recent measurements of $\Mt$ is
observed.  The point with the arrow indicates the prediction when only
the photon vacuum polarisation is included in the electroweak
radiative corrections, which shows that the precision electroweak
Z-pole data are sensitive to non-trivial electroweak corrections.  The
error due to the uncertainty on $\alpha(\MZ^2)$ (shown as the length
of the arrow) is not much smaller than the experimental error on
$\swsqeffl$ from {\LEPI} and SLD.  This underlines the continued
importance of a precise measurement of
$\sigma(\mathrm{e^+e^-\rightarrow hadrons})$ at low centre-of-mass
energies.

Of the measurements listed in Table~\ref{tab-SMIN}, $\Rl$ is one of
the most sensitive to QCD corrections.  With $\MZ=91.1875$~\GeV{}, and
imposing $\Mt=173.3\pm1.1$~\GeV{} as a constraint,
$\alfas=0.1223\pm0.0038$ is obtained.  Alternatively,
$\slept\equiv\shad/\Rl=2.0003\pm0.0027~\nb$~\cite{bib-Z-pole} which
has higher sensitivity to QCD corrections and less dependence on $\MH$
yields: $\alfas=0.1179\pm0.0030$.  The typical errors arising from the
variation of $\MH$ between $100~\GeV$ and $200~\GeV$ are of the order
of $0.001$, and are somewhat smaller for $\slept$.  These results on
$\alfas$, as well as those reported in the next section, are in very
good agreement with both the world average, $\alfmz=0.1184 \pm
0.0007$\cite{common_bib:pdg2010}, and the value $\alfmz=0.1178 \pm
0.0033$ which is based solely on NNLO QCD results excluding the
{\LEPI} lineshape results and accounting for correlated
errors~\cite{QCD:Bethke:2000, *Bethke:2004uy}.

\clearpage

\section{Standard Model Analyses}

In the following, several different SM analyses as reported in
Table~\ref{tab-BIGFIT} are discussed.  The $\chi^2$ minimisation is
performed with the program MINUIT~\cite{MINUIT}, and the predictions
are calculated with ZFITTER~6.43 as a function of the five SM input
parameters $\dalhad$, $\alfmz$, $\MZ$, $\Mt$ and $\LOGMH$ which are
varied simultaneously in the fits. The fit procedure is described in
detail in Reference~\citen{bib-Z-pole}. The somewhat large
$\chi^2$/d.o.f.{} for all of these fits is caused by the large
dispersion in the values of the leptonic effective electroweak mixing
angle measured through the various asymmetries at {\LEPI} and
SLD~\cite{bib-Z-pole}.  Following~\cite{bib-Z-pole}, this dispersion
is interpreted as a fluctuation in one or more of the input
measurements, and thus we neither modify nor exclude any of them.  A
further significant increase in $\chi^2$/d.o.f.{} is observed when the
NuTeV results are included in the analysis.

\begin{figure}[htbp]
\begin{center}
\includegraphics[width=0.9\linewidth]{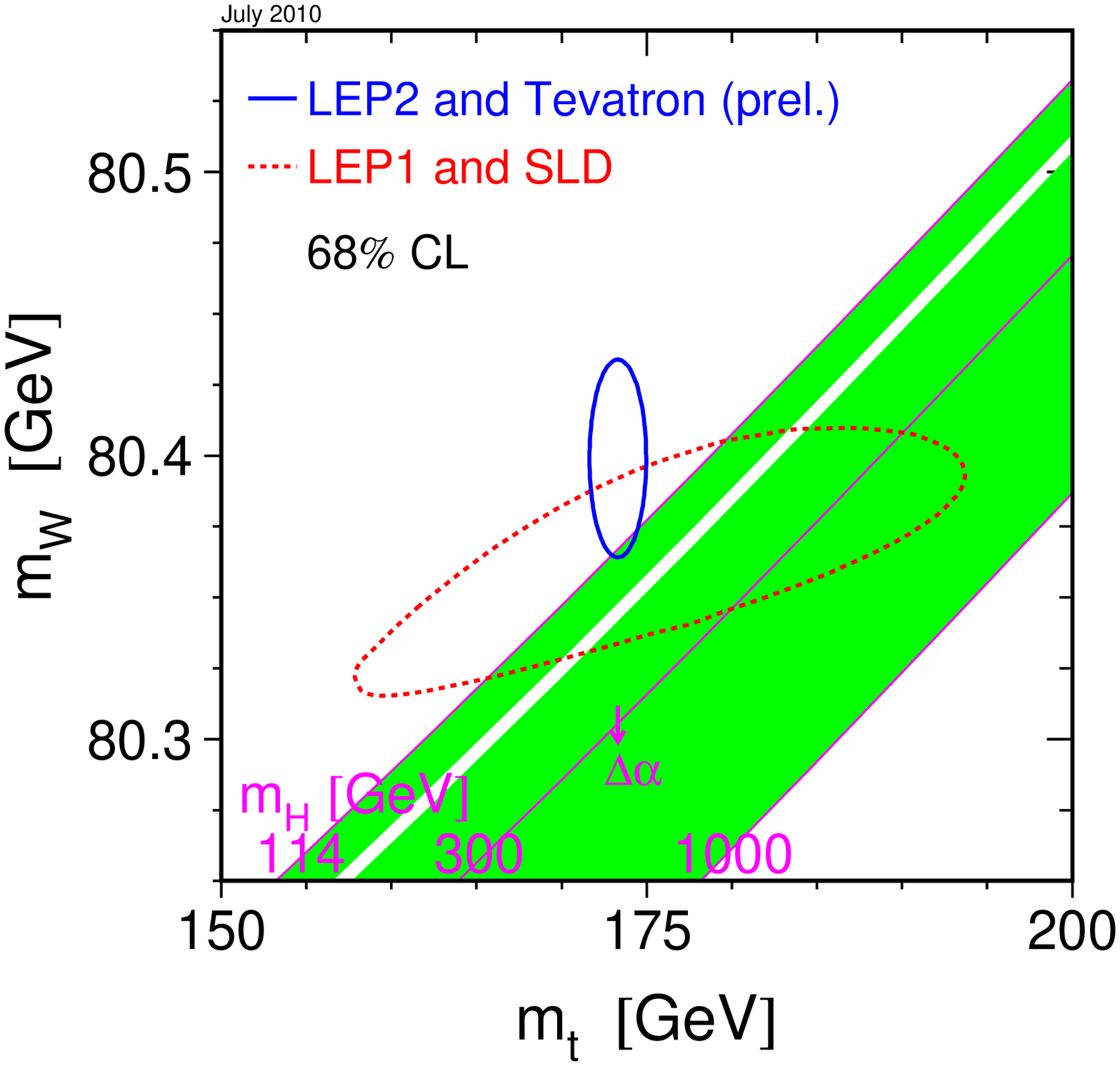}
\caption[]{ The comparison of the indirect constraints on $\MW$ and
  $\Mt$ based on $\LEPI$/SLD data (dashed contour) and the direct
  measurements from the $\LEPII$/Tevatron experiments (solid contour).
  In both cases the 68\% CL contours are plotted.  Also shown is the
  SM relationship for the masses as a function of the Higgs mass in
  the region favoured by theory ($<1000~\GeV$) and not excluded by
  direct searches ($114~\GeV$ to $158~\GeV$ and $>175~\GeV$).  The
  arrow labelled $\Delta\alpha$ shows the variation of this relation
  if $\alpha(\MZ^2)$ is varied between $-1$ and $+1$ standard
  deviation. This variation gives an additional uncertainty to the SM
  band shown in the figure.}
\label{fig:mtmW}
\end{center}
\end{figure}

To test the agreement between the Z-pole data~\cite{bib-Z-pole}
({\LEPI} and SLD) and the SM, a fit to these data is performed.  The
result is shown in Table~\ref{tab-BIGFIT}, column~1. The indirect
constraints on $\MW$ and $\Mt$ are shown in Figure~\ref{fig:mtmW},
compared with the direct measurements.  Also shown are the SM
predictions for Higgs masses between 114 and 1000~\GeV.  As can be
seen in the figure, the indirect and direct measurements of $\MW$ and
$\Mt$ are in good agreement, and both sets prefer a low value of the
Higgs mass.

\begin{figure}[htbp]
\begin{center}
\includegraphics[width=0.9\linewidth]{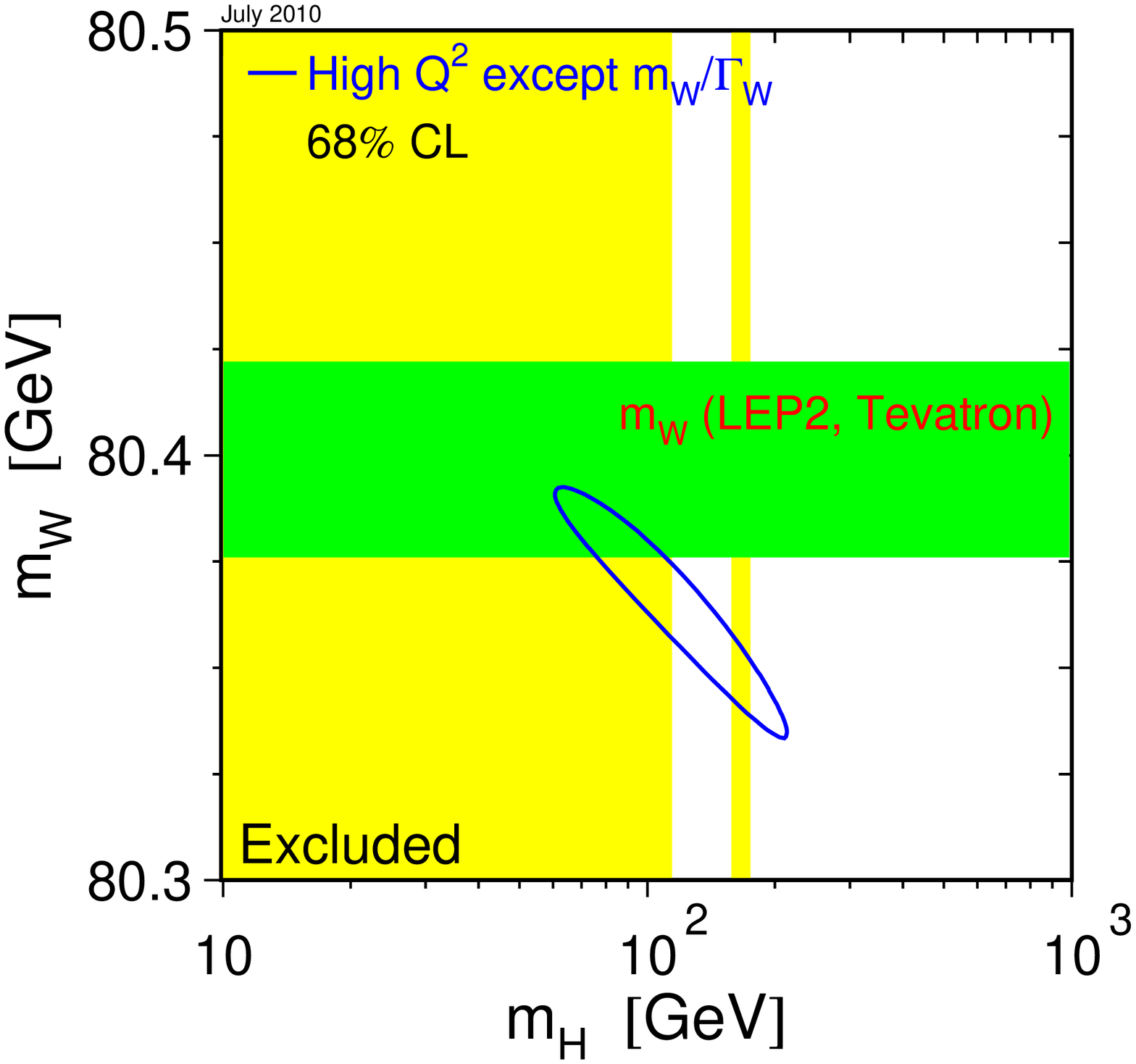}
\end{center}
\vspace*{-0.6cm}
\caption[]{ The 68\% confidence level contour in $\MW$ and $\MH$ for
  the fit to all high-$Q^2$ data except the direct measurement of
  $\MW$, indicated by the shaded horizontal band of $\pm1$ sigma
  width.  The vertical bands shows the 95\% CL exclusion limit on
  $\MH$ from the direct searches at LEP-II (up to $114~\GeV$) and the
  Tevatron ($158~\GeV$ to $175~\GeV$).  }
\label{fig-mhmw}
\end{figure}

\begin{figure}[htbp]
\begin{center}
\includegraphics[width=0.9\linewidth]{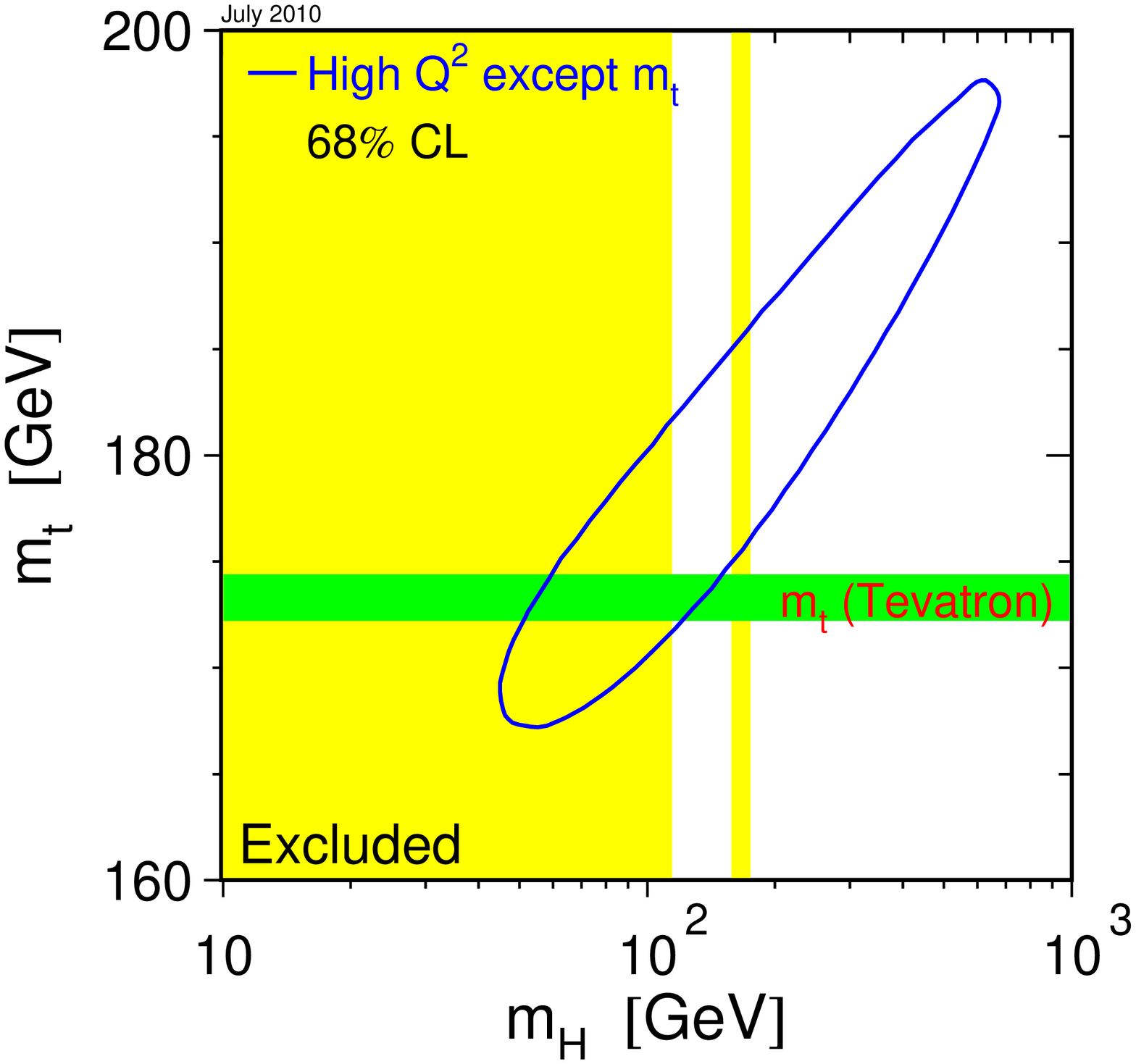}
\end{center}
\vspace*{-0.6cm}
\caption[]{ The 68\% confidence level contour in $\Mt$ and $\MH$ for
  the fit to all high-$Q^2$ data except the direct measurement of
  $\Mt$, indicated by the shaded horizontal band of $\pm1$ sigma
  width.  The vertical band shows the 95\% CL exclusion limit on $\MH$
  from the direct searches at LEP-II (up to $114~\GeV$) and the
  Tevatron ($158~\GeV$ to $175~\GeV$).  }
\label{fig-mhmt}
\end{figure}

For the fit shown in column~2 of Table~\ref{tab-BIGFIT}, the direct
$\Mt$ measurement is included to obtain the most precise indirect
determination of $\MW$.  The result is also shown in
Figure~\ref{fig-mhmw}.  Also in this case, the indirect determination
of the W boson mass, $80.365\pm0.020~\GeV$, is in agreement with the
direct measurements from {\LEPII} and the Tevatron, $\MW=
80.399\pm0.023~\GeV$.  For the fit shown in column~3 of
Table~\ref{tab-BIGFIT} and Figure~\ref{fig-mhmt}, the direct $\MW$ and
$\GW$ measurements from {\LEPII} and the Tevatron are included instead
of the direct $\Mt$ measurement in order to obtain the most precise
prediction, $\Mt= 179^{+12}_{-9}~\GeV$, in good agreement with the
direct measurement of $\Mt = 173.3\pm1.1~\GeV$.

\begin{figure}[htbp]
\begin{center}
\includegraphics[width=0.9\linewidth]{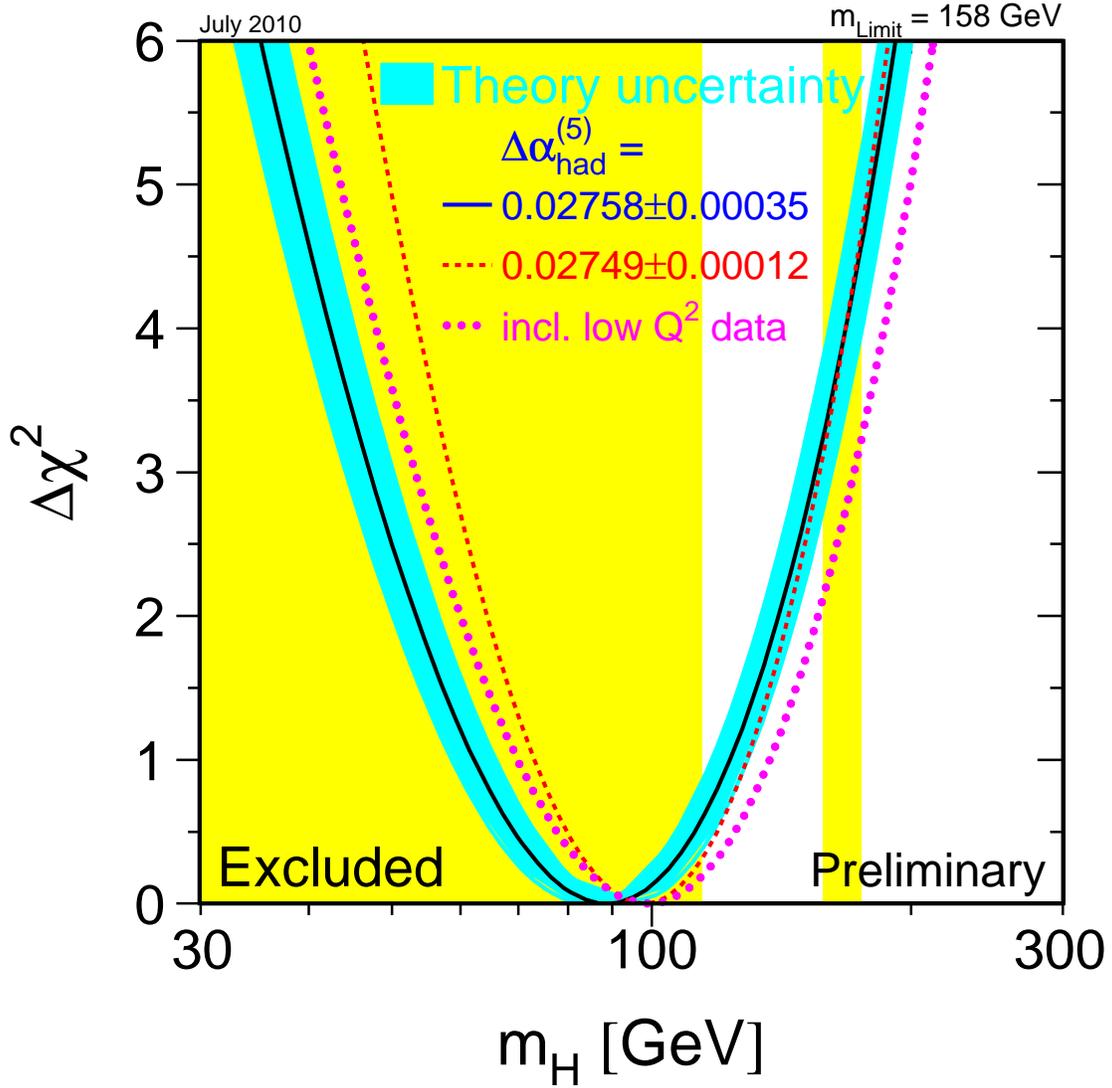}
\end{center}
\vspace*{-0.6cm}
\caption[]{ $\Delta\chi^{2}=\chi^2-\chi^2_{min}$ {\it vs.} $\MH$
  curve.  The line is the result of the fit using all high-$Q^2$ data
  (last column of Table~\protect\ref{tab-BIGFIT}); the band represents
  an estimate of the theoretical error due to missing higher order
  corrections.  The vertical band shows the 95\% CL exclusion limit on
  $\MH$ from the direct searches at LEP-II (up to $114~\GeV$) and the
  Tevatron ($158~\GeV$ to $175~\GeV$). The dashed curve is the result
  obtained using the evaluation of $\dalhad$ from
  Reference~\citen{bib-Troconiz-Yndurain-2004}. The dotted curve
  corresponds to a fit including also the low-$Q^2$ data from
  Table~\ref{tab-SMpred}. }
\label{fig-chiex}
\end{figure}

Finally, the most stringent constraints on $\MH$ are obtained when all
high-$Q^2$ measurements are used in the fit.  The results of this fit
are shown in column~4 of Table~\ref{tab-BIGFIT}.  The predictions of
this fit for observables measured in high-$Q^2$ and low-$Q^2$
reactions are listed in Tables~\ref{tab-SMIN} and~\ref{tab-SMpred},
respectively.  In Figure~\ref{fig-chiex} the observed value of
$\Delta\chi^2 \equiv \chi^2 - \chi^2_{\mathrm{min}}$ for this fit
including all high-$Q^2$ results is plotted as a function of $\MH$.
The solid curve is the result using ZFITTER, and corresponds to the
last column of Table~\ref{tab-BIGFIT}.  The shaded band represents the
uncertainty due to uncalculated higher-order corrections, as estimated
by ZFITTER.

The 95\% one-sided confidence level upper limit on $\MH$ (taking the
theory-uncertainty band into account) is $158~\GeV$.  The 95\%
C.L. lower limit on $\MH$ of 114.4~\GeV{} obtained from direct
searches at LEP-II\cite{LEPSMHIGGS} and the region between $158~\GeV$
and $175~\GeV$ excluded by the Tevatron
experiments~\cite{TeVNPHWG:2010} are not used in the determination of
this limit.  Including the LEP-II direct-search limit increases the
limit from $158~\GeV$ to $185~\GeV$.  Also shown is the result (dashed
curve) obtained when using $\dalhad$ of
Reference~\citen{bib-Troconiz-Yndurain-2004}.

\begin{table}[htbp]
\renewcommand{\arraystretch}{1.5}
  \begin{center}
\begin{tabular}{|c||c|c|c|c|c|}
\hline
&     - 1 -              &      - 2 -             &    - 3 -               &     - 4 -             \\
& all Z-pole             & all Z-pole data        & all Z-pole data        & all Z-pole data       \\[-3mm]
& data                   &    plus   $\Mt$        & plus $\MW$, $\GW$      & plus $\Mt,\MW,\GW$    \\
\hline
\hline
$\Mt$\hfill[\GeV] 
& $173^{+13 }_{-10}$     & $173.3^{+1.1}_{-1.1}$  & $179^{+12}_{- 9}$      & $173.4^{+1.1}_{-1.1}$  \\
$\MH$\hfill[\GeV] 
& $111^{+190}_{-60}$     & $117^{+58}_{-40}$      & $146^{+241}_{-81}$     & $ 89^{+35}_{-26}$      \\
$\log_{10}(\MH/\GeV)$  
& $2.05^{+0.43}_{-0.34}$ & $2.07^{+0.18}_{-0.19}$ & $2.16^{+0.42}_{-0.35}$ & $1.95^{+0.14}_{-0.15}$ \\
$\alfmz$          
& $0.1190\pm 0.0027$     & $0.1190\pm0.0027$      & $0.1190\pm 0.0028$     & $0.1185\pm 0.0026$     \\
\hline
$\chi^2$/d.o.f.{} ($P$)
& $16.0/10~(9.9\%)$      & $16.0/11~(14\%)$       & $16.9/12~(15\%)$       & $17.3/13~(19\%)$       \\
\hline
\hline
$\swsqeffl$
& $\pz0.23149$           & $\pz0.23149$           & $\pz0.23143$           & $\pz0.23138$ \\[-1mm]
& $\pm0.00016$           & $\pm0.00016$           & $\pm0.00014$           & $\pm0.00013$ \\
$\swsq$     
& $\pz0.22331$           & $\pz0.22328$           & $\pz0.22287$           & $\pz0.22301$ \\[-1mm]
& $\pm0.00062$           & $\pm0.00040$           & $\pm0.00036$           & $\pm0.00028$ \\
$\MW$\hfill[\GeV]
& $80.363\pm0.032$       & $80.365\pm0.020$       & $80.386\pm0.018$       & $80.379\pm0.015$   \\
\hline
\end{tabular}
\end{center}
\caption[]{ Results of the fits to: (1) all Z-pole data ({\LEPI} and
  SLD), (2) all Z-pole data plus direct $\Mt$ determination, (3) all
  Z-pole data plus direct $\MW$ and $\GW$ determinations, (4) all
  Z-pole data plus direct $\Mt,\MW,\GW$ determinations (i.e., all
  high-$Q^2$ results).  As the sensitivity to $\MH$ is logarithmic,
  both $\MH$ as well as $\log_{10}(\MH/\GeV)$ are quoted.  The bottom part
  of the table lists derived results for $\swsqeffl$, $\swsq$ and
  $\MW$.  See text for a discussion of theoretical errors not included
  in the errors above.  }
\label{tab-BIGFIT}
\renewcommand{\arraystretch}{1.0}
\end{table}

\begin{table}[htbp]
\begin{center}
  \renewcommand{\arraystretch}{1.30}
\begin{tabular}{|ll||r||r|r|l|}
\hline
 && {Measurement with}  & {Standard Model} & {Pull}  \\
 && {Total Error}       & {High-$Q^2$ Fit} & {    }  \\
\hline
\hline
&APV~\cite{QWCs:theo:2003:new}
                &                        &                      &       \\
\hline
&$\QWCs$        & $-72.74\pm0.46\pzz\pz$ & $-72.911\pm0.029\pzz$& $0.4$ \\
\hline
\hline
&M{\o}ller~\cite{E158RunI, *E158RunI+II+III}
                &                        &                      &        \\
\hline
&$\swsqMSb$     & $0.2330\pm0.0015\pz$   & $0.23110\pm0.00013$  & $1.3$  \\
\hline
\hline
&$\nu$N~\cite{bib-NuTeV-final}
                &                        &                      &        \\
\hline
&$\gnlq^2$      & $0.30005\pm0.00137$    & $0.30399\pm0.00016$  & $2.9$  \\
&$\gnrq^2$      & $0.03076\pm0.00110$    & $0.03012\pm0.00003$  & $0.6$  \\
\hline
\end{tabular}\end{center}
\caption[]{ Summary of measurements performed in low-$Q^2$ reactions,
  namely atomic parity violation, $e^-e^-$ M{\o}ller scattering and
  neutrino-nucleon scattering.  The SM results and the pulls
  (difference between measurement and fit in units of the total
  measurement error) are derived from the SM fit including all
  high-$Q^2$ data (Table~\ref{tab-BIGFIT}, column~4) with the Higgs
  mass treated as a free parameter.}
\label{tab-SMpred}
\end{table}

\begin{figure}[p]
\vspace*{-1.0cm}
\begin{center}
\includegraphics[height=20cm]{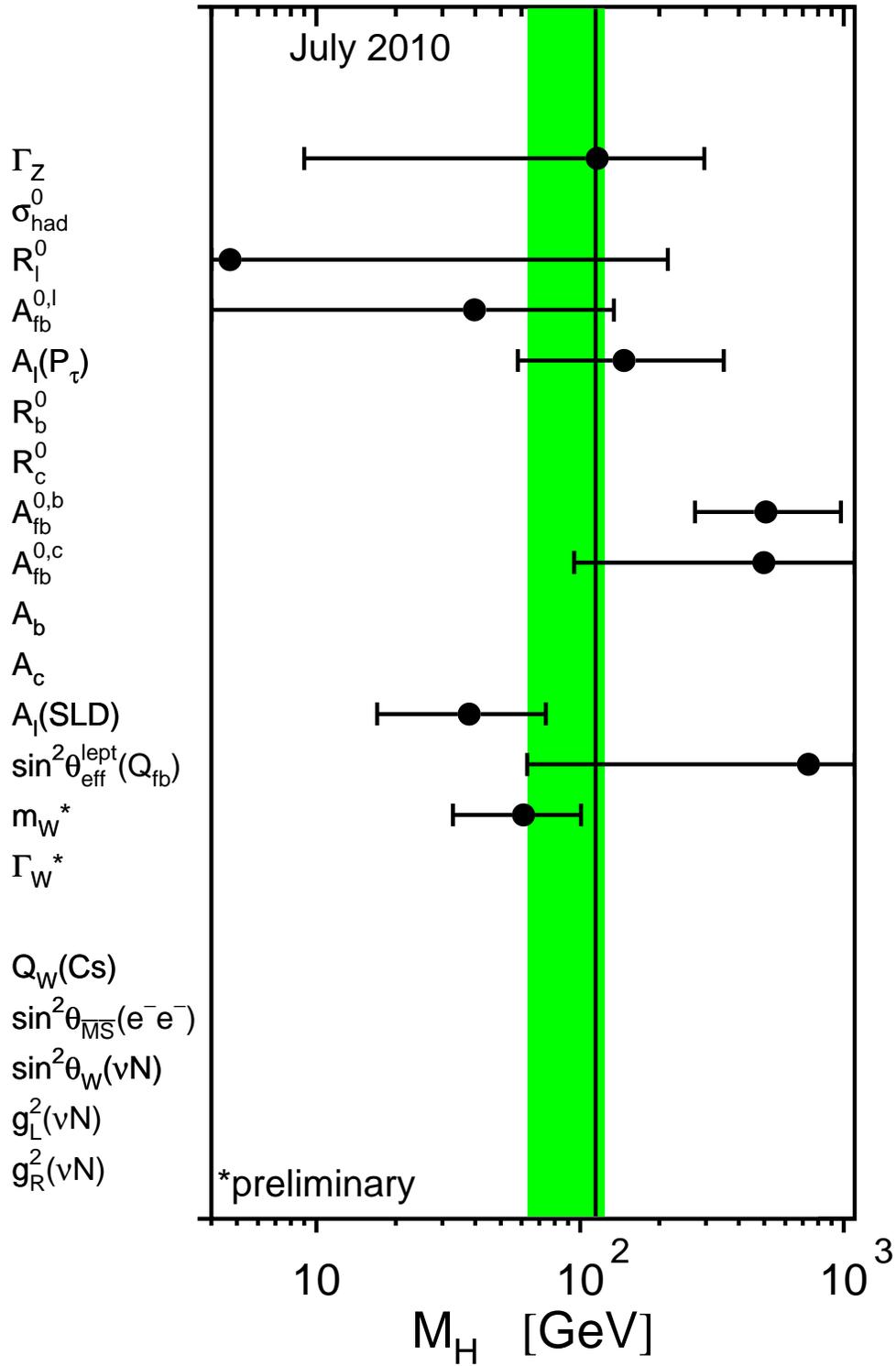}
\end{center}
\vspace*{-0.6cm}
\caption[]{ Constraints on the mass of the Higgs boson from each
  pseudo-observable. The Higgs-boson mass and its 68\% CL uncertainty
  is obtained from a five-parameter SM fit to the observable,
  constraining $\dalhad=0.02758\pm0.00035$, $\alfmz=0.118\pm0.003$,
  $\MZ=91.1875\pm0.0021~\GeV$ and $\Mt=173.3\pm1.1~\GeV$.  Because of
  these four common constraints the resulting Higgs-boson mass values
  are highly correlated.  The shaded band denotes the overall
  constraint on the mass of the Higgs boson derived from all
  pseudo-observables including the above four SM parameters as
  reported in the last column of Table~\ref{tab-BIGFIT}. The vertical
  line denotes the 95\% CL lower limit from the direct search for the
  Higgs boson. Results are only shown for observables whose
  measurement accuracy allows to constrain the Higgs-boson mass on the
  scale of the figure. }
\label{fig-higgs-obs}
\end{figure}

Given the constraints on the other four SM input parameters, each
observable is equivalent to a constraint on the mass of the SM Higgs
boson. The constraints on the mass of the SM Higgs boson resulting
from each observable are compared in Figure~\ref{fig-higgs-obs}.  For
very low Higgs-masses, these constraints are qualitative only as the
effects of real Higgs-strahlung, neither included in the experimental
analyses nor in the SM calculations of expectations, may then become
sizeable~\cite{Kawamoto:2004pi}.  Besides the measurement of the W
mass, the most sensitive measurements are the asymmetries, \ie,
$\swsqeffl$.  A reduced uncertainty for the value of $\alpha(\MZ^2)$
would therefore result in an improved constraint on $\log\MH$ and thus
$\MH$, as already shown in Figures~\ref{fig-gllsef} and
\ref{fig-chiex}.


\clearpage

\boldmath
\section{Conclusions}
\label{sec-Conc}
\unboldmath

The preliminary and published results from the LEP, SLD and Tevatron
experiments, and their combinations, represent tests of the Standard
Model (SM) at the highest interaction energies.  The combination of
the numerous precise electroweak results
yields stringent constraints on the SM and its free parameters.  Most
measurements agree well with the predictions.  The spread in values of
the various determinations of the effective electroweak mixing angle
in asymmetry measurements at the Z pole is somewhat larger than
expected~\cite{bib-Z-pole}.  
\boldmath
\section{Prospects for the Future}
\unboldmath

The measurements from data taken at or near the Z resonance, both at
LEP as well as at SLC, are final and published~\cite{bib-Z-pole}.
Some improvements in accuracy are expected in the high energy data
(\LEPII), where each experiment has accumulated about 700~pb$^{-1}$ of
data, when combinations of the published final results are made. The
measurements from the Tevatron experiments will continue to improve
with the increasing data samples collected during Run~II.  The
measurements of $\MW$ have almost reached a precision comparable to
the uncertainty on the prediction obtained via the radiative
corrections of the Z-pole data and the top-quark mass, providing an
important test of the Standard Model. The large data samples to be
collected at the LHC set the stage for further improvements.  Work is
needed in reconciling the definition of the top-quark mass in the
theory and its extraction from the collider data.

\section*{Acknowledgements}

We would like to thank the accelerator divisions of CERN, SLAC and
Fermilab for the efficient operations and excellent performance of the
LEP, SLC and Tevatron accelerators. We would also like to thank
members of the
E-158 and NuTeV collaborations for useful discussions concerning their
results.  Finally, the results on constraints on the Standard Model
parameters would not be possible without the close collaboration of
many theorists.

\clearpage

\bibliographystyle{PhysRep}
\bibliography{s10_ew,common,gg,ff,smat,fsi,be,4f_s06,gc,mw}

\end{document}